\documentclass[letterpaper]{jpconf}
\usepackage{graphicx}
 \DeclareGraphicsExtensions{.eps,.ps,.eps.gz,.ps.gz}  

\begin{document}
\title{Exclusive Physics at the Tevatron}
\author{James Pinfold}

\address{University of Alberta}

\begin{abstract}
Groundbreaking studies of exclusive physics processes at the Tevatron 
have  demonstrated that protonÐ(anti)proton colliders are
 not only quarkÐantiquark but also  gluonÐgluon colliders,  photonÐphoton and 
 photon - gluon colliders. These studies are briefly reviewed in this paper.
\end{abstract}

\section{Introduction}
\medskip

\begin{figure}[htbp]
  \begin{center}
    \includegraphics[width=12.0cm]{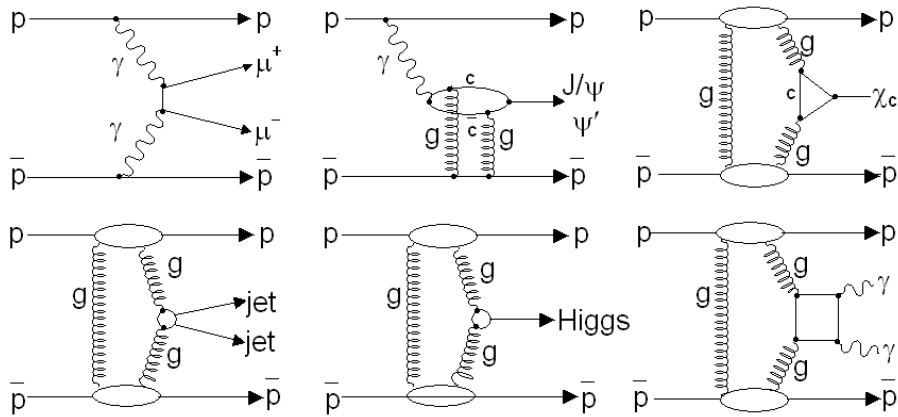}
    \caption{The three basic diagrams explored by the exclusive physics effort at the Tevatron are shown above: QED two-photon, interaction (top left); photoproduction, or photon-pomeron fusion (top middle and top right); and,  gluon-gluon fusion with a third ÒscreeningÓ  gluon (bottom). This interaction is also termed double pomeron exchange. Only at the LHC will  the mass of the central state  will be large enough to allow the Higgs boson  to be produced. \label{feynmans}}
  \end{center}
\end{figure}

Most reactions at Fermilab's Tevatron occur when the colliding proton and antiproton break apart into quarks and gluons that hadronize to form the particles that fly off into the detector. In exclusive interactions, however, the proton and antiproton avoid the breakup, glancing off each other in a process where the underlying interaction involves some combination of photons and/or gluons. The basic Feynman diagrams that describe exclusive processes  at the Tevatron are shown in Figure~\ref{feynmans}. These diagrams describe exclusive processes generated by photon - photon, gluon - photon and gluon - gluon interactions.

\section{The Processes $\gamma\gamma \rightarrow \e^{+}\e^{-}$ and $g g  \rightarrow \gamma\gamma$ }
\medskip
In 2006, the CDF collaboration at the Tevatron obtained the first clear evidence for exclusive interactions at a hadron collider, when they observed high-energy photon pairs in the central rapidity (barrel) region of the detector, but with nothing else, down to an angle of around 0.1$^{o}$  from the beam ($\pm$7.4 units of pseudorapidity). They found only three events initially, against a small background predicted to be at most 0.2 events \cite{CDF2007}. These events were consistent with being produced via gluon - gluon ÒfusionÓ via a quark ÒboxÓ where the gluons originate from the beam particles, as shown in  Figure~\ref{feynmans}. An additional ÒscreeningÓ gluon is exchanged to cancel the colour of the interacting gluons and allow the leading hadrons to stay intact. The collaboration has since observed more exclusive two-photon final state candidates in new data.

The search for this unusual two-photon process at the Tevatron was originally proposed in 2001, when we  first explored the possibility that the Higgs boson could be produced by gluon-gluon fusion as described in Figure~\ref{feynmans} \cite{ALBROW2005}. The idea is that if the Higgs field fills the vacuum, it should be possible to Òexcite the vacuumÓ into a real Higgs particle in a glancing collision of a proton and antiproton. Theorists had various estimates for the probability of this happening, but their predictions varied widely.

The two-photon process we measured in the CDF detector is produced in much the same way as the Higgs would be, but much more prolifically, so making it a Òstandard candleÓ for the production of Higgs bosons. Theorists from the Centre for Particle Theory at the University of Durham predicted that there should be only about one clean two-photon event of this kind in data corresponding to 532 pb$^{-1}$ of integrated luminosity taken by CDF in Run II at the Tevatron \cite{KHOZE2006}. The three events that the CDF collaboration found confirmed this prediction. Thus, the similar process that could produce the elusive Higgs boson must also happen, and could be measured at the LHC, thereby providing measurements of the particleÕs mass, spin and other properties.

In the process of checking this measurement, we  came across another exclusive physics process that had never been seen before at a protonÐ(anti)proton collider. We found 16 events that are consistent with the QED prediction that photons travelling with the proton and antiproton can interact to produce only an electron-positron pair ($\gamma\gamma \rightarrow e^{+}e^{Ð}$) without breaking up the proton and antiproton \cite{CDF2007A}. In this case the Tevatron acts as a photon ÒcolliderÓ. As the backgrounds to this process are similar to the final state discussed above, we gained further confidence in our exclusive two-photon final state analysis. To date, we have found many more exclusive electron-positron candidate events. QED two-photon processes such as this, which have previously been observed in electron-positron, electron-proton and nuclear collisions, should provide a means of calibrating the momentum scale and resolution of forward proton spectrometers proposed for the ATLAS and CMS experiments at the LHC \cite{FP420}.

\section{Exclusive Charmonia and Bottomonia}
\medskip
We then reasoned that we should also see exclusive muon-pair events produced by the same QED interaction, as in Figure~\ref{feynmans}. Apart from an indication of exclusive pair production at the ISR at CERN \cite{ANTREASYAN1980}, this would be another ÒfirstÓ at a protonÐ(anti)proton collider. In 2007 our supposition was confirmed, but with an added bonus. The expected process, $\gamma\gamma \rightarrow  \mu^{+}\mu^{Ð}$, was indeed detected according to QED expectations.

In addition, the CDF physicists recorded, for the first time in hadron-hadron collisions, exclusive photoproduction of the J/$\psi$ and $\psi$(2S) decaying to a pair of muons (Figure~\ref{feynmans}). The team also detected the contribution from exclusive production via gluon-gluon fusion of the $\chi_{c0}$, decaying to a muon pair and a soft photon (Figure~\ref{feynmans}). Evidence for this state in CDF data had also been reported earlier, in 2003 \cite{WYATT2003}. The cross sections we measured are, d$\sigma$/dy $|_{y=0}$  for J/$\psi$, $\psi$(2S) and $\chi_{c0}$ are 3.92 $\pm$ 0.25(stat) $\pm$ 0.52(syst) nb, 0.53 $\pm$ 0.09(stat) $\pm$ 0.10(syst) nb, and 76 $\pm$ 10(stat) $\pm$ 10(syst) nb, respectively, and the continuum $\mu^{+}\mu^{-}$ production is consistent with QED predictions. The differential cross-section measured for $\chi_{c0}$  production is consistent with the predictions of the Durham Group \cite{DURHAMCHIC}. More details of the study of exclusive production of charmonia are available elsewhere \cite{CHARMONIAPRL}. 

An analysis aimed at higher muon-pair masses including the Upsilon ($\Upsilon$) region
-  with 8 $ < M(\mu^{+}\mu^{-}) < $ 40 GeV/c$^{2}$ - is underway. The QED continuum
in this region  is a good control, and the photoproduced $\Upsilon$ states have cross sections 
that are within reach, although not very well known. (The HERA data \cite{HERAUPSILONS}  do not resolve clearly 
the $\Upsilon$(1S), $\Upsilon$(2S) and $\Upsilon$(3S) states and have quite large uncertainties).  Predictions for
[ d$\sigma/dy]_{y=0}$ are around 5-14 pb \cite{UPSILONXS}. Applying the branching fraction to 
$\mu^{+}\mu^{-}$, B = 0.025, would give a few hundred events with 2 fb$^{-1}$ ($\times$ the acceptance and efficiency). 
 In CDF, with y = 0,  we have W(J/$\psi$) $\sim$80 GeV and W($\Upsilon$(1)) $\sim$136 GeV. 
 At HERA the ratio of these cross sections is $\sim$ 300.  For  dimuons we used a trigger with two muons 
 with p$_{T} (\mu)  > $ 4 GeV/c and $|\eta| < $ 0.6. The inclusive muon pair mass  spectrum shows the 
 states $\Upsilon$(1S), $\Upsilon$(2S) and $\Upsilon$(3S) as well-separated
peaks,  see e.g. Ref. \cite{UPSILONMASSPEAK}.

We can move towards  a definition of  ``exclusivity''  by requiring no other tracks on the
$ \mu^{+}\mu^{-}$  vertex, $(\pi - \Delta\phi) < $  0.1 rad and p$_{T}(\mu^{+}\mu^{-})  <  $ 1.5 GeV/c.  These
cuts should be efficient for the QED events and for most of the $\Upsilon$'s; the issue is non-exclusive
backgrounds. We are now studying the continuum to see
if it is exactly as expected for QED, as a control, and can then give $\Upsilon$-photoproduction cross
sections, depending on unknown backgrounds from $\chi_{b} \rightarrow  \Upsilon + \gamma$. 
Unfortunately, neither the $\chi_{b}$ production rates nor their radiative decays are well known. 
The p$_{T }(\Upsilon)$  distribution is broader for $\chi_{b}$-daughters, which may help, but one 
would like to reconstruct the photons, which is probably
not possible with pile-up. The same issues will confront us at the LHC.

\section{Exclusive Z Production}

Exclusive Z photoproduction is allowed in the Standard Model.  However the SM cross 
section is too small  for excluisve Z production to be seen at the Tevatron 
\cite{EXCLUSIVEZTEV}. At the LHC (14 TeV) 
the predictions are as high as  69 fb \cite{EXCLUSIVEZTEV}, which may make an observation possible. We used 
a sample of 3.17 $\times$ 10$^{5}$ lepton pairs with mass greater than 40 GeV/c$^{2}$ , 
of which 1.83 $\times$ 10$^{5}$ were in the Z peak. We required exclusivity over the full range 
$ |\eta| <$ 7.4,  finding  8 events, agreeing  with QED expectations.  All events were very 
back-to-back in the transverse plane, with ($\pi - \Delta\phi$) $<$  0.75$^{o}$. One event with mass equal
to  66 GeV/c$^{2}$ had an antiproton track in the Roman pots; for the others the antiproton  was 
out of their acceptance or they were not operational. None of the eight exclusive events were 
Z -candidates, and a limit was placed: $\sigma_{exc}$(Z ) $<$ 0.96 pb at 95\% C.L. \cite{EXCLUSIVEZCDF} 
 It may be possible to improve this limit using  a factor 2-3 more data and including pile-up events, 
 using the requirements of no associated  tracks on the $ l^{+}l^{-}$  vertex, small ($\pi - \Delta\pi)$ and 
 p$_{T} (l^{+} l ^{-}$). We are testing this method on the  Upsilon region, with 8 $<$ M ($l^{+}l^{-})$  $<$ 40 GeV/c$^{2}$ .

\section{Exclusive Di-jet Production}
\medskip
After publishing results on exclusive lepton-pair and photon-pair production, the CDF collaboration scored a hat-trick in 2008 when it published results on exclusive di-jet production, as in Figure~\ref{feynmans}  \cite{DIJETPUB}. Using a Roman Pot deployed tracker some 66 m from the interaction point to tag the unbroken antiproton in conjunction with a large rapidity gap on the deflected proton side, the team defined a sample of potentially exclusive events. The greater the share of the mass of the central system that the two jets enjoyed, the Òmore exclusiveÓ the events were expected to be. This expectation was borne out by the Monte Carlo simulation \cite{MONKPILKINGTON} for central exclusive production and in agreement with the predictions of the Durham Group \cite{KHOZE2007}. As the di-jet fractional share of the overall central mass of the event tended to one - and the exclusive di-jet sample became purer and purer  - evidence for b-jet suppression was seen, as theoretically expected. As in the case of exclusive $\gamma\gamma$ and $\chi_{c0}$ production, this is an example of the Tevatron acting as a gluon-gluon collider. The detection at the Tevatron of these exclusive processes, resulting from gluon-gluon interactions, strongly suggests that exclusive production of the Higgs boson by the similar process would be detected at the LHC.

\section{Conclusion}
\medskip
Although forward proton detectors have been used to study Standard Model physics for a couple of decades, the new landscape revealed by exclusive physics at hadron colliders has been fully realized only in the past few years. In this arena, the LHC is not only preparing to take the baton from the Tevatron, but also to enter the race with greatly improved tools. The FP420 R\&D project \cite{FP420} is planning to provide the means to measure the displacement and angle of the outgoing protons from exclusive interactions by deploying high precision ``edgeless'' silicon trackers less than a centimetre from the beam, at $\pm$420 m from the beam intersection points of the ATLAS and CMS experiments at the LHC. This gives these experiments the ability to calculate the proton momentum loss and transverse momentum, allowing the mass of the centrally produced system to be reconstructed with a resolution of a few GeV/c$^{2}$ per event whatever the central system. Broadly speaking then, in the exclusive physics arena, the LHC becomes a Òmulti-colliderÓ, where the gluon-gluon, photon-photon, or photon-pomeron ``beam energy'' is known.

The ability of the FP420 detectors to measure intact protons from an exclusive interaction, in conjunction with the associated centrally produced system using the current ATLAS and/or CMS detector, will provide rich new perspectives at the LHC on studies in QCD, electroweak physics, the Higgs sector and beyond Standard Model physics. In some scenarios, these detectors may be the primary means of discovering new particles at the LHC, with unique ability to measure their quantum numbers. The addition of the FP420 detectors will thus, for a relatively small cost, significantly enhance the discovery and physics potential of the ATLAS and CMS experiments. The existence proof provided by the exclusive physics results from the Tevatron shows that such a programme is feasible.

\section*{References}
\bibliography{iopart-num}

\end{document}